\documentclass[preprint2]{aastex}
\usepackage{epsfig}
\newif\ifAMStwofonts
\AMStwofontstrue

\def\mpc{\rm{h^{-1}Mpc}}

\def\gsim{~\rlap{$>$}{\lower 1.0ex\hbox{$\sim$}}}

\def\fesc{f_{\rm esc}}

\def\ltsim{\lower.5ex\hbox{$\; \buildrel < \over \sim \;$}}
\def\gtsim{\lower.5ex\hbox{$\; \buildrel > \over \sim \;$}}
\def\ltsim{\lower.5ex\hbox{$\; \buildrel < \over \sim \;$}}
\def\gtsim{\lower.5ex\hbox{$\; \buildrel > \over \sim \;$}}

\def\vx{{\bf x}}

\def\vr{{\bf r}}

\newcommand{\op}{Ly$\alpha$\ }
\newcommand{\ob}{Ly$\beta$\ }

\begin{document}
\title{ STATISTICS OF NEUTRAL REGIONS DURING HYDROGEN RE-IONIZATION}

\author {Adi Nusser\altaffilmark{1}, Andrew J. Benson\altaffilmark{2},
Naoshi Sugiyama\altaffilmark{3,4}, Cedric Lacey\altaffilmark{5,6}}
\altaffiltext{1}{Physics Department and Space Science Institute- Technion, 
Haifa 32000, Israel,
 {adi@physics.technion.ac.il}}
\altaffiltext{2}{California Institute of Technology, 
MC 105-24, 1200 E. California Blvd., Pasadena, CA 91125, U.S.A.,
abenson@astro.caltech.edu} 
\altaffiltext{3}{Division of Theoretical Astrophysics, National
Astronomical Observatory Japan, Mitaka, 
181-8588,~Japan, naoshi@yso.mtk.nao.ac.jp} 
\altaffiltext{4}{Max-Planck-Institut fur Astrophysik, Karl-Schwarzschild-Str. 1, Postfach 1317
D-85741 Garching, Germany}
\altaffiltext{5}{SISSA, via Beirut 2-4, 34014 Trieste, Italy}
\altaffiltext{6}{Department of Physics, University of Durham, UK}
     

\begin{abstract}

We present predictions for two statistical measures of the hydrogen
reionization process at high redshift. The first statistic is the
number of neutral segments identified in spectra of high redshift QSOs
as a function of their length. The second is the cross-correlation of
neutral regions with possible sources of ionizing radiation.  These
independent probes are sensitive to the topology of the ionized
regions. If reionization proceeded from high to low density regions
then the cross-correlation will be negative, while if voids were ionized
first then we expect a positive correlation and a relatively small
number of long neutral segments. We test the sensitivity of these
statistics for reionization by stars in high redshift galaxies.  The
flux of ionizing radiation emitted from stars is estimated by
identifying galaxies in an N-body simulation using a semi-analytic
galaxy formation model.  The spatial distribution of ionized gas is
traced in various models for the propagation of the ionization fronts.
A model with ionization proceeding from high to low density regions is
consistent with the observations of Becker et al. (2001), while models
in which ionization begins in the lowest density regions appear to be
inconsistent with the present data.
\end{abstract}


\section{Introduction}

Early hydrogen reionization is a particularly interesting process in
the high redshift universe and is inevitably linked to the appearance
of the first star forming objects, at least those that served as
sources of the ionizing radiation.  If it occurred early enough,
reionization imprints distinct features in maps of the cosmic
microwave background (CMB) on arcminute angular scales
\citep{vishniac87,bruscoli, BNSL}. Several aspects of hydrogen
reionization remain uncertain despite the rapidly accumulating data on
the high redshift universe. For example, it is unclear what objects
produce most of the ionizing radiation, although high redshift
galaxies are very strong candidates \citep{CR86,haiman96,ciardi99}.
It is also unclear how the ionized regions develop in space
\citep{miralda}.  The ionizing sources are likely to lie in high
density regions, but those regions do not necessarily ionize first;
the ionizing photons may tunnel into less dense regions and ionize
those first.  Further, the duration of reionization is unknown and
only a lower limit on the redshift marking the end of that epoch
exists \citep{becker,gunn65}.

In previous papers (Benson et al. 2001, Liu et al. 2001) we examined
how the CMB is affected by the reionization process and how future CMB
maps can be used to extract information on that process.  However,
hydrogen reionization is currently best probed by spectra of high
redshift QSOs. Unlike maps of the CMB which are sensitive to line of
sight integrals over the density and velocity of ionized gas, QSO
spectra contain direct information on the local distribution of
neutral hydrogen.

Recently, Becker et al. (2001) analyzed spectra of a sample of QSOs
with redshifts between $z=5.82$ and $z=6.28$. In the spectrum of their
highest redshift QSO ($z=6.28$), the transmitted flux in the \op\ and
\ob\ forest in the redshift stretch $5.95<z<6.16$ is consistent with
zero, with a lower limit of 20 on the \op optical depth. This long
stretch in redshift corresponds to a comoving distance of $60\mpc$ in
a universe with a cosmological constant of $\Lambda_0=0.7$ and matter
density of $\Omega_0=0.3$. Becker et al. suggest that this long
neutral region is a detection of the end of the hydrogen reionization
era.  To increase the signal-to-noise ratio, Becker et al. binned
their spectra in $4$\AA\ pixels. This prevented them from detecting
small scale dark windows that are likely to appear in the spectra as
left overs from the reionization epoch.  A high resolution spectrum for
one of the quasars observed by Becker et al. was obtained by
Djorgovski et al. (2001). The spectrum of this quasar ($z=5.73$) was
thorough analyzed by Djorgovski et al. (2001) and was found to contain
several small dark windows signifying the detection of the trailing
edge of the reionization epoch.

Motivated by the results of Becker et al. (2001) and Djorgovski et
al. (2001), we examine here how the key ingredients in the
reionization process can be probed by QSO spectra and future high
redshift galaxy surveys.  The methodology of the present paper has
been previously developed in Benson et al. (2001, hereafter BNSL). As
in BNSL, we obtain the distribution of ionized gas in an N-body
simulation.  Using a semi-analytic model for galaxy formation
\citep{kauff93,somerville99,coleetal99} we identify mock galaxies in
the simulation and estimate the ionizing flux emitted by stars at high
redshift.  We then use several schemes to follow the development of
ionized regions in the simulation.  Using the output of this procedure
we obtain predictions for two statistics. The first is the expected
number of neutral segments longer than a given length, and the second
is the cross-correlation function between the galaxies and neutral
regions. Both of these measures rely on observations of QSO spectra,
and the latter also on a sample of high redshift candidates for the
sources of ionizing radiation.

\section{Modeling the development of ionized regions}

BNSL employed a semi-analytical model for galaxy formation in a high
resolution N-body simulation of dark matter to estimate the amount of
ionizing radiation produced by stars in high redshift galaxies.  Here
we follow a similar procedure using the latest version of the {\sc
galform} galaxy formation model \citep{coleetal99}, and the same
$\Lambda$CDM simulation as BNSL (c.f.  Jenkins et al. 1998). This
simulation has $\Omega _0 = 0.3$, a cosmological constant $\Lambda _0
= 0.7$, a Hubble constant of $h=0.7$ in units of $100{\rm km
s^{-1}Mpc^{-1}}$, and is normalized to produce the observed abundance of
rich clusters at $z\approx 0$ according to Eke, Cole \& Frenk (1996).
The simulation has a box of length 141.3 $\rm h^{-1}Mpc$ and contains
$256^3$ dark matter particles.

Most of the ionizing photons emitted by stars are likely to be
absorbed by gas and dust inside galaxies and only a small fraction,
$\fesc$, escapes and becomes available for hydrogen ionization in the
intergalactic medium (IGM) \citep{leitherer95}.  Assuming a
value\footnote{BNSL considered several models for the variation of
$f_{\rm esc}$ from galaxy to galaxy. Here we adopt the simplest model,
in which $f_{\rm esc}$ is constant for all galaxies.} for $\fesc$,
BNSL used the following models to follow the propagation of ionized
regions in the simulation (see BSNL for details).

{\it Model A (Growing front model)} Ionize a spherical volume around
each source (halo) with a radius equal to the ionization front radius
for that halo assuming a large-scale uniform distribution of neutral
hydrogen.  Since the neutral hydrogen in the simulation is not
uniformly distributed, and also because some spheres will overlap, the
ionized volume will not contain the correct total mass of 
 hydrogen. We therefore scale the radius of each sphere by a
constant factor and keep repeating the procedure until the correct
total mass has been ionized.

{\it Model B (High density model)} We simply rank the cells in the
simulation volume by their density. We then completely ionize the gas
in the densest cell. If this has not ionized enough hydrogen we ionize
the second densest cell. This process is repeated until the correct
total mass of hydrogen has been ionized.

{\it Model C (Low density model)} As model B, but we begin by ionizing
the least dense cell, and work our way up to cells of greater and
greater density \citep{miralda}.

{\it Model D (Random spheres model)} As Model A but the spheres are
placed in the simulation entirely at random rather than on the dark
matter halos. 

{\it Model E (Boundary model)} Ionize a spherical region around each
halo with a radius equal to the ionization front radius for that
halo. This may ionize too much or not enough neutral hydrogen
depending on the density of gas around each source. We therefore begin
adding or removing cells at random from the boundaries of the already
ionized regions until the required mass is ionized.

Guided by the observations of Becker et al. (2001) we will compute the
number of neutral segments and the cross correlations at three output
redshifts, $z=6.67$, $6.22$, and $5.80$.  Table 1 lists the volume
filling factors (ratio of volume of ionized regions to total volume of
the simulation box) in each of our five models for $\fesc=0.01$
(column 2 in table 1) and $\fesc=0.05$ (column 3).  We will see later
that the results of Becker et al. imply that the amount of ionizing
radiation increases significantly between $z\approx 6.2$ and $5.8$.
Therefore we also show results for a variable escape
fraction,$\fesc^{\rm var}$ (column 4) which equals $0.01$ before
$z=6.22$ and increases linearly with time to $0.1$ at $z=5.8$.  As
expected, model C (low density model) has the highest filling factor
for a given $\fesc$.  For $\fesc=0.05$ the simulation box is fully
ionized at $z=5.8$ in all models.

\begin{table}
\caption{The volume filling factor of ionized regions in the
simulation at $z=6.66$, 6.22, and 5.80, for two constant values of
$\fesc$ (columns 3 and 4), and a variable fraction $\fesc^{\rm var}$
(column 4) that increases linearly with time from 0.01 at $z=6.22$ to
0.1 at $z=5.80$.}
\begin{tabular}{|l|c|c|c|} \hline
model &$\fesc$ &$\fesc$&$\fesc{^{\rm var}}$ \\
 &$0.01$ &$0.05$&$0.1$\\ \hline 
A\quad$z=6.66$&$0.155$&$ 0.612$&\\
A\quad$z=6.22$&$0.187$&$0.892$&\\
A\quad$z=5.80$&$0.210$&$1$&0.835 \\ \hline
B\quad$z=6.66$&$0.079$&$0.464 $&\\ 
B\quad$z=6.22$&$0.101$&$0.688$&\\
B\quad$z=5.80$&$0.114$&$ 1 $&0.624\\ \hline
C\quad$z=6.66$&$0.458$&$0.892$& \\ 
C\quad$z=6.22$&$0.536$&$1$& \\
C\quad$z=5.80$&$0.589$&$ 1 $&0.982 \\ \hline
D\quad$z=6.66$&$0.229$&$0.726$&\\ 
D\quad$z=6.22$&$0.270$&$0.919$&\\ 
D\quad$z=5.80$&$0.315$&$ 1 $&0.922 \\ \hline
E\quad$z=6.66$&$0.183$&$0.648$&\\ 
E\quad$z=6.22$&$0.217$&$0.937$& \\
E\quad$z=5.80$&$0.238$&$ 1$ &0.871\\ \hline
\end{tabular}
\label{table1}
\end{table}
\section{The statistical measures}

\subsection{\it The number of neutral segments}

\begin{figure}
\vspace{1.cm}
\hspace{7.cm}
\mbox{\epsfig{figure=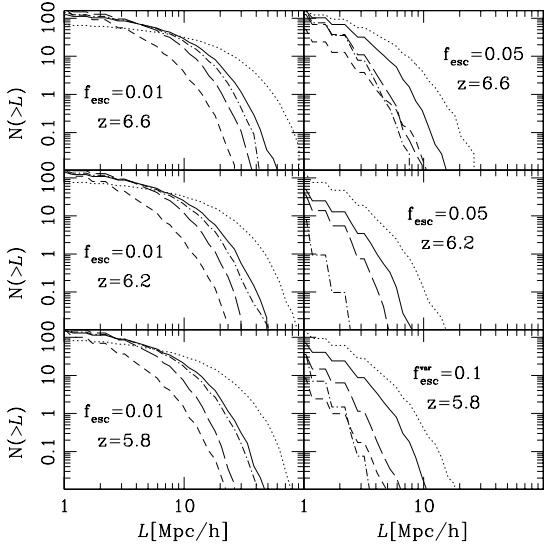,height=7.5cm,angle=180}}
\vspace{-2.0cm}
\caption{The mean  number of neutral segments of length $>L$  in
lines of sight each of redshift span corresponding to $1{\rm h^{-1}
Gpc}$, as estimated from the simulation.  The lines in each panel
refer to the five ionization models: solid, dotted, short dashed, long
dashed, and, dotted-short dashed lines correspond to models A, B, C,
D, and E, respectively. There are no neutral regions in model C with
$\fesc=0.05$ at $z=6.22$.  Models with $\fesc^{\rm var}$ had
$\fesc=0.01$ for $z>6.22$, and increasing linearly to
$\fesc=0.05$ from $z=6.22$ to $z=5.80$.}
\label{figure1}
\end{figure}

We are now in a position to compute the proposed statistics.  We begin
with $N(>L)$, the mean number of neutral (unionized) segments of length
greater than $L$ in a given redshift range in a line of sight (see
Barkana 2002, for a similar statistic).  We have the spatial
distribution of ionized and neutral regions in the simulation as a
function of redshift, for each of our five reionization scenarios (see
Table 1).  To compute $N(>L)$ we choose several random ``lines of
sight'' in the output of the simulation at a given redshift. In each
line of sight we identify the neutral segments and tabulate their
lengths, $L$, in comoving $\mpc$.  A line of sight is obtained by
starting from a grid point at the boundary of the simulation and going
around the boundary of a rectangular slice of perimeter $4 \times
141{\rm h^{-1}Mpc}$ (comoving) until we return to the starting
point. (The shape of the path used to extract a line of sight makes no
difference to our results. Using a rectangular path helps reduce the
chance of pattern repetition.) This yields a total of $256$ lines of
sight each spanning a redshift range corresponding to $564 {\rm
h^{-1}Mpc}$ (comoving).  We then compute the mean $N(>L)$ from these
lines of sight.  For convenience we normalize $N(>L)$ to a redshift
span of $1 {\rm h^{-1}Gpc}$ by multiplying the direct result obtained
from the simulation by $1000/564$.  The $N(>L)$ (normalized to $1 {\rm
h^{-1}Gpc}$) is shown in figure (\ref{figure1}) for $z=6.66$ (top),
6.22 (middle), and 5.8 (bottom).  The panels to the left show $N(>L)$
computed for $\fesc=0.01$, at these three redshifts.  To the right we
show curves computed with $\fesc=0.05$ at $z=6.66$ (top) and $6.22$
(middle), and a variable $\fesc^{\rm var}$ at $z=5.8$ (bottom).  We
have also computed $N(>L)$ from lines of sight each of length
$141{\rm h^{-1} Mpc}$ passing through in the simulation box at random
positions and found very similar results to those shown in the figure.

In observed spectra ionized regions with large optical depths can be
confused with completely unionized regions, and vice versa.  By
measuring \ob absorption lower limits of about 20 on the \op optical
depth can be obtained. In CDM-like models ionized regions with optical
depth larger than this lower limit are small. A careful analysis and
modeling of spectra can therefore help reduce this confusion.
Nevertheless to estimate the degree of the confusion, we have computed
$N(>L)$ assuming that regions with optical depth larger than 20 are
identified as unionized\footnote{The neutral hydrogen fraction at each
point was estimated assuming photoionization 
equilibrium (Theuns et al. 1998) and a  hydrogen ionization
rate of $\Gamma=4.3\times 10^{-13}$s$^{-1}$. The 
\op\ line-width and  effects of peculiar velocities were also taken
into account.}. We found that in this more detailed calculation $N(>L)$ is
shifted to smaller $L$ by less than a factor of 2 at large $L$. At
small $L$, $N(>L)$ is reduced by a factor of less than 2, although
this could likely be reduced with a more careful analysis as suggested
above.

The cells of our computational grid are approximately $0.55h^{-1}$Mpc
(comoving) in extent. According to Benson et al. (2002) the
characteristic smoothing length in the IGM at this redshift is
approximately $0.2h^{-1}$Mpc (comoving). Thus, our current simulation
does not fully resolve the structure of gas in the IGM. To assess the
consequences of this limitation we repeated our calculations using a
lower resolution computational grid ($128^3$ cells instead of
$256^3$). We find that, at large $L$, the lengths of neutral regions
are approximately 30\% smaller when the higher resolution grid is
used. Doubling the grid resolution to $512^3$ cells (and therefore
almost fully resolving the smoothing length) should make little
difference to our results. We also checked that noise due to the
finite number of particles is unimportant for the results presented
here.

\subsection{\it The cross correlation}
In addition to the distribution of neutral hydrogen, the
cross-correlation, $\xi$, requires the spatial distribution of
(potential) sources of the ionizing photons. In real observations the
distribution of neutral hydrogen is obtained from QSO spectra and the
positions of sources from a high redshift galaxy or QSO survey.  Our
goal here is simply to demonstrate that $\xi$ can distinguish between
various models for the propagation of the ionized regions.  We will
therefore simply compute $\xi$ from the three dimensional distribution
of neutral hydrogen in the simulation, i.e. we will not address the
question of how well $\xi$ can be estimated from realistic mock
observations where the neutral regions are identified in lines of
sight.

The simulation box is divided into a $256^3$ cubic grid.  At each grid
point $\vx$ we define a quantity $h$ to be zero if that point has been
ionized and unity otherwise.  We also use the cloud-in-cell (CIC)
method to derive the number density $n(\vx)$ of galaxies from the
galaxy positions in the simulation. We select galaxies on the basis of
their ionizing luminosity and include only those with luminosities
sufficiently high to ensure the population is fully resolved in the
N-body simulation. Because ionizing luminosity correlates only weakly
with halo mass---since it depends so strongly on the star formation
rate---this means we select only the most luminous sources---$6$, $6$,
and $10\times10^{54}h^{-2}$photons/s at $z=5.8$, $6.2$ and $6.7$
respectively. These sources are  rare and contribute only
a small fraction to the total ionizing luminosity density of the
universe (about 15\% and 18\% at z=6.222 and 5.80, respectively).      
Denoting the average values of $n$ and $h$ by $\bar n$ and
$\bar h$ respectively, we define the cross-correlation, $\xi$, as
\begin{equation}
\xi(\vr)=<[n(\vx)/\bar n -1][h(\vx+\vr)/ \bar h -1]>
\end{equation}
where the symbol $<.>$ implies averaging over all grid points $\vx$.
In practice the calculation of $\xi$ is done using the technique of
Fast Fourier Transforms. The results are shown in Figure 2.  This
figure demonstrates that $\xi$ is sensitive to the ionization model.
It is positive for model C (low density), almost vanishes for model
D (random spheres), and is negative for models A, B, and E, which
ionize dense regions first.  However, $\xi$ has a similar shape for
all the high density models (A, B and E) and we expect that it will
be difficult to distinguish between them in observational data.
Nevertheless they all are significantly different from either model C
or D. So $\xi$ should successfully discriminate among low density,
random ionization, and high density models.

\begin{figure}
\vspace{1.cm}
\hspace{7.cm}
{\epsfig{figure=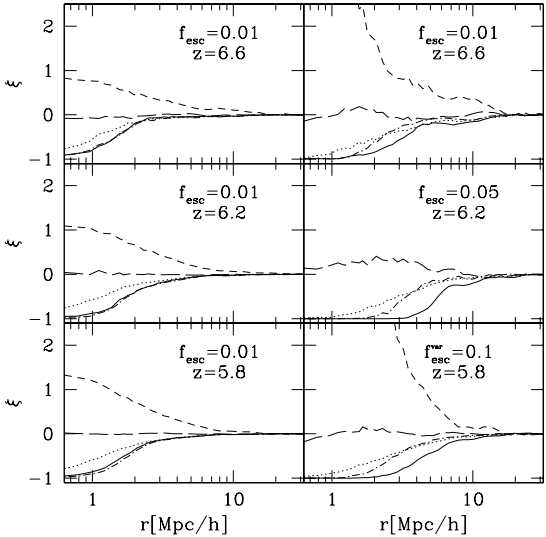,height=7.5cm,angle=180}}
\vspace{-2.0cm}
\caption{The cross-correlations between the galaxy distribution and
neutral regions in the simulation.  The notation of the lines is the
same as in the previous figure.}
\label{corrhi}
\end{figure}

\section{Discussion}

The statistic $N(>L)$ giving the number of neutral segments of length
greater than $L$ for a given total length of a QSO spectrum is
sensitive to the filling factor and the way in which ionization
proceeds.  The cross correlation between candidate ionizing sources
and neutral regions is less sensitive to the filling factor but is a
more direct and robust probe of the propagation of the ionization
fronts. In addition to QSO spectra, the cross-correlation requires a
sample of candidates (galaxies and QSO) for the ionizing radiation.
Catalogs of galaxies and QSO at high redshift are rapidly accumulating,
making it possible to compute the QSO-flux and galaxy-flux
correlations. Comparison between galaxy-flux and QSO-flux
correlation functions  will tell us
whether galaxies or QSO  contributed most of the ionizing radiation.

Current observations do not allow a robust determination of $N(>L)$,
The number of spectra needed to determine $N(>L)$ to within a given
accuracy can be estimated by noting that the relative error on this
function is $1/\sqrt{M N}$, where $M$ is the number of observed QSO
spectra covering the same redshift range.  Nevertheless we still can
make general conclusions based on the Becker et. al.  (2001) result,
assuming that the long Gunn-Peterson trough they observe is indeed a
signature of reionization.  Let us take the length of a spectrum in the
\op forest at
$z\approx 6$ to be $\sim 250\mpc$ corresponding  to the comoving distance
between \op and \ob emission lines.  An
inspection of Figure \ref{figure1} shows that: $1)$ A completely
neutral stretch of a comoving length of $60\mpc$ at $z\approx 6.2$ is
inconsistent with a large filling factor. 
2) The observations indicate that the chances of finding long neutral
regions at $z<5.94$ are tiny while they are significant at higher
redshift.  This behavior seems inconsistent with our theoretical
$N(>L)$ computed with constant $\fesc$. If $\fesc=0.01$ then there are
similar probabilities for finding long segments at $z=6.22$ and
$z=5.80$.  If $\fesc=0.05$ then the box is fully ionized at $5.80$,
while at $z=6.22$ long segments are very rare.  Therefore the data
favor models in which there is a significant increase in the amount of
ionizing radiation in the IGM, either due to an increasing escape
fraction over this redshift range, or a much stronger evolution in the
galaxy/QSO population than is predicted by our model. And, $3)$ a
model in which ionization proceeds from low to high density regions
seems to be inconsistent with a $\sim 60 \mpc$ neutral region even for
escape fractions as low as $\fesc=0.01$.
\section*{Acknowledgments} 
This research is supported by the EC RTN network ``The Physics of the
Intergalactic medium''. A.N. is supported by the Israeli Academy of
Science, the German Israeli Foundation for Scientific Research and
Development, and the Technion V. P. R. Fund and Henri Gutwirth
Promotion of Research.  N.S. is supported by the Alexander von
Humboldt Foundation and a Japanese Grant-in-Aid for Science Research Fund of
the Ministry of Education, No. 14540290.  CGL is supported by PPARC. 
A.N. and A.J.B. wish to
thank the National Astronomical Observatory of Japan in Mitaka for its
hospitality and support.

\end{document}